\documentclass[12pt]{article}

\usepackage{amsmath}
\usepackage{graphicx}
\usepackage{subfig}
\usepackage{amssymb}

\begin{document}

\title{Modeling of neuron-semiconductor interactions in neuronal networks interfaced with silicon chips}
\author{Nikesh S. Dattani}
\date{May 28, 2008}

\maketitle

\begin{center}
Department of Applied Mathematics and Department of Biology \\ University of Waterloo \\ Waterloo, Ontario~
 N2L 3G1, Canada
\end{center}

\begin{abstract}
  Recent developments in the interfacing of neurons with silicon chips may pave the way for progress in constructing scalable neurocomputers.  The assembly of synthetic neuronal networks with predefined synaptic connections and controlled geometric structure has been realized experimentally within the last decade.  Furthermore, when such neuronal networks are interfaced with semiconductors, action potentials in neurons of the network can be elicited by capacitative stimulators, and voltage measurements can be made by transistors incorporated into the associated silicon chip.  Despite the impressive progress, such preliminary devices have not yet demonstrated the performance of useful computations, and constructing larger devices can be both expensive and time-consuming.  Accordingly, an appropriate modeling framework with the capability to simulate current experimental results in such devices may be used to make useful predictions regarding their potential computational power.  A proposed modeling framework for functional neuronal networks interfaced with silicon chips is presented below.
\end{abstract}

\newpage

\tableofcontents
\newpage

\section{Introduction}

\qquad There are two motives that are currently driving the desire to construct a neurocomputer.  Firstly, attempting to discover the mechanisms by which vertebrate brains demonstrate intelligence is extremely difficult when studying such structures \textit{in vivo}, due to their enormous complexity and due to our shortage in non-invasive procedures for probing such systems with single-cell resolution\cite{Bulloch1992}.  Although cultured reconstructions of neuronal networks have provided us with more optimism, such assemblies remain structurally complex, and their synaptic wiring remains unclear\cite{2005Merz}.  The construction of simple neuronal networks with \textit{predefined} synaptic wiring will improve studies on the functionality of such systems.  

Secondly, conventional computers are reaching their limits.  The search for a more powerful computational paradigm has lasted for more than a decade, and the physical realization of a scalable quantum computer has been proven to be exceptionally challenging\cite{Dattani2008}.  A technology capable of exploiting neurons to perform computations may address this problem, since it is well-recognized that neurons have incredible potential. 

Recent developments have been made towards the construction of neurocomputers
\cite{2000Prinz}\cite{2005Merz}; however, these constructions are still very simple, and it can be expensive and time-consuming to construct synthetic neuronal networks that are adequately large to perform interesting computations.  The primary objective of this study is to develop a mathematical model that describes the neurodynamics in existing attempts to implement synthetic neuronal networks with functionality.  Once such a mathematical model is able to reproduce experimental observations in the existing \textit{simple} synthetic neuronal networks (such as voltage readings of action potentials), the model can then extrapolate to predict the behaviour of more complex networks.  Consequently, we can perform computational experiments in neurocomputing, and this may serve as a step towards the investigation of the practicality of such computational devices.  


\section{Review of Literature}

\subsection{Neuronal Networks in Culture}

\subsubsection{Historical Notes}

\qquad Some of the simplest nervous systems present in the animal kingdom are those of the phyla Cnidaria\cite{1999Taddei-Ferretti} and Echinodermata.  Such systems (which are called ``nerve nets'') may be relatively easy to study, but are not capable of many of the higher-order functions in which theoretical neuroscientists are often interested.  Adding to the challenge, invertebrate nervous systems capable of executing such functions are often too complex to study as a whole, and accordingly, methods have been developed to reconstruct partial circuits of identified neurons in cell cultures.

  Such practice began in 1972 with the isolation of single nerve cells from the brains of \textit{Lymnaea stagnalis} molluscs (freshwater snails) which were then cultivated \textit{in vitro} by Kostenko\cite{1972Kostenko}.  A few years later, in 1979, a group of researchers at the California Institute of Technology described neuronal networks of the marine snail \textit{Aplysia californica} in cell culture\cite{1979Kaczmarek}.  Such work soon became more customary and was adopted by a great deal of neuroscientists in the early 1980s[7-11].
  
  
  \subsubsection{Significant Progressive Developments}

\qquad In 1990, a study involving the \textit{in vitro} reconstruction of the respiratory central pattern generator in \textit{Lymnaea stagnalis} made a discovery regarding the mechanism of its function that would not otherwise have been made if the system was not studied in cell culture\cite{1990Syed}.  Based on previous studies of the intact nervous system, it was suspected that the generation of the respiratory rhythm driving expiratory and inspiratory lung movements in the snail was enforced by some subset of three particular neurons (but which combination of these three were actually required, and whether or not more neurons were required still remained a mystery).  These three cells were then isolated and grown in cell culture.  It was discovered that the appropriate rhythmic activity was not evoked in any of the individual neurons when isolated.  Likewise, no two \textit{pairs} of these neurons elicited the necessary activity, but when \textit{all three} of the neurons were interconnected, the system exhibited a rhythmic output identical to the activity observed in the intact nervous system.  In this fashion, studying a reconstructed neuronal network \textit{in vitro} lead to a discovery that one was not able to conclude at the time, based on studies of the entire intact nervous system.   

\subsubsection{Remaining Challenges}

\qquad As incredible as the above discovery was, more complicated neurobiological functions involving larger and much more tortuous networks of neurons may be impractical to study using this technique, since the synaptic wiring often still remains unclear in cell cultures.  This lead to the development of neuronal networks in which we have predefined the synaptic wiring.

\subsection{Neuronal Networks with Predefined Topology}

\subsubsection{Historical Notes}

\qquad Guided outgrowth of neurites has been accomplished by several research groups since 1985[13-19], but the formation of synapses under explicit control was not realized until more recently.


\subsubsection{Significant Progressive Developments}

\qquad In January of 2000, Astrid Prinz and Peter Fromherz of the Max Planck Institute for Biochemistry implemented and reported a procedure for controlling synapse formation by guided outgrowth of neurites in neurons from \textit{Lymnaea stagnalis}.  

On the culture dish, proteinaceous conditioning factors obtained from the snail's brain were adsorbed to a substrate coated with polylysine.  The conditioning factors were then locally inactivated in particular regions by ultraviolet illumination, which resulted in defined geometric patterns of conditioning factors which would eventually act as guides for the outgrowth of the neurites.  These patterns were manifested in the form of linear strips 14$\mu m$ in width.  Pairs of \textit{Lymnaea} neurons were then placed at opposite ends along these strips and separated by several hundred micrometers.  Over time, neurites would grow from these neurons, following the strips of conditioning factors, and synaptogenesis would occur as the growth cones of the two neurons collided (as displayed in Fig.1).      

\begin{figure}[h!]
	\centering
	\includegraphics[width = \textwidth] {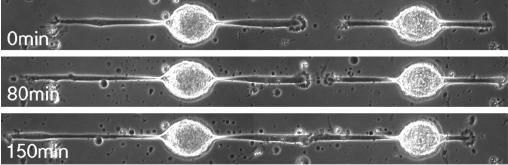} 
	\caption{Guided encounter of the growth cones of two neurons from \textit{Lymnaea stagnalis}.  The width of the linear trace of proteinaceous conditioning factors is 14$\mu m$.  This figure was reproduced with the kind permission of the publisher and authors of Ref. \cite{2000Prinz}.}
\end{figure}

Action potentials in a presynaptic neuron were induced by injecting a depolarizing current, and this lead to action potentials in the corresponding postsynaptic neuron in cases where there was strong enough coupling between the two neurons.  This method can then be extended to formulate networks of neurons with higher populations and with more connections.

\subsubsection{Remaining Challenges}

\qquad A crucial shortcoming of guiding the growth of neurites in this manner emerges from the fact that the neurons often cannot maintain their adhesion to the surfaces for extended culture times\cite{1994Fromherz,2004Nam}.  Additionally, one needs a reliable method to stimulate particular neurons, and to measure the responses of members of the network at the single-cell level.  Both of these shortcomings can be addressed by neuronal networks interfaced with silicon ships.  
  
\subsection{Interfacing Neurons with Silicon Chips}

\qquad An alternative to directing the growth of neurites by chemical tracks on a culture dish involves crafting pre-defined topographical structures on the surface of silicon chips.  Such topographical structures include circular pits connected by 14$\mu m$ wide grooves (as displayed in Fig. 2a).

\begin{figure}[h!]
  \centering
  \subfloat[Circular pit and grooves on silicon surface]{\includegraphics[width=0.47\textwidth,height=1.3in]{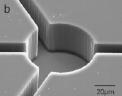}}~~~~
  \subfloat[Neuronal network interfaced with the chip]{\includegraphics[width=0.47\textwidth,height=1.3in]{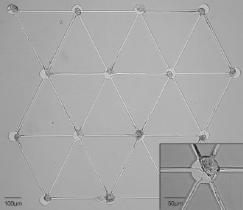}} 
  \caption{A silicon chip interfaced with a topographically controlled network of neurons.  This figure was reproduced with the kind permission of the publisher and authors of Ref. \cite{2005Merz}.  Copyright Wiley-VCH Verlag GmbH \& Co. KGaA. }
\end{figure}   

	Additionally, the bottom of each pit contains a capacitative stimulator for inducing depolarizations (and desirably, action potentials), and a transistor used to probe the responses of the neurons.  These microstructures are not visible in Fig.2.  As in the case of the pairs of interconnected neurons whose neurites were guided by chemical tracts, stimulations made using the capacitative stimulator are able to induce action potentials in individual neurons, which are often able to elicit action potentials in a proceeding neuron.

  Such experiments provide a fresh feel of optimism towards the construction of synthetic neuronal networks with functionality, which will be a central step in the journey towards building a scalable neurocomputer.  However, no mathematical modeling of the electrodynamics associated with such constructions has been published thus far.  The following chapter describes how I propose to model this system, and how I anticipate to utilize the model to describe the reported experimental results, as well as to make valuable predictions for future experiments in neurocomputing.	

\section{Computational Modeling}

\subsection{Electrophysiological Modeling}

\qquad The first task to confront is to distinguish an appropriate modeling framework for the electrophysiological properties of the particular neurons which I am considering in this work: \textit{Lymnaea stagnalis}.

\subsubsection{The Moris-Lecar Model}

  To begin with, let us consider the Morris-Lecar (ML) model\cite{1981Morris} for generating action potentials in single-cells:

\vspace{-2mm}
\begin{equation}
C\frac{dV}{dt} ~=~ -g_{Ca} m_{\infty} (V)(V - V_{Ca}) - g_K w (V - V_K) - g_L (V - V_L) + I_{app}
\end{equation}

\vspace{-2mm}
\begin{equation}
~~\frac{dw}{dt} = \frac{w_{\infty}(V) - w}{\tau(V)}, 
\end{equation}

where

\vspace{-2mm}
\begin{equation}
m_{\infty}(V) = 0.5 \left[ 1 + \textrm{tanh} \left( \frac{V-v_1}{v_2} \right) \right ] 
\end{equation}

\vspace{-2mm}
\begin{equation}
w_{\infty}(V) = 0.5 \left[ 1 + \textrm{tanh} \left( \frac{V-v_3}{v_4} \right) \right ]
\end{equation}

\vspace{-2mm}
\begin{equation}
~ \tau (V) = \frac{1}{\phi \, \textrm{cosh}\left(  \frac{V - v_3}{2 v_4} \right)}
\end{equation}

A complete list describing the identities of the above symbols used is provided in the appendix.  This model is able to provide a reasonable description of the voltage-dynamics associated with action potentials in neurons \textit{in vivo}.  Particularly, it uses an instantaneously responding voltage-sensitive Ca$^{+2}$ conductance for excitation and a delayed voltage-dependent K$^+$ conductance for recovery\cite{2007Lecar}.  For this reason, it will be necessary to modify this model to account for the fact that the action potentials in the neurons of these synthetic networks are induced electrically by capacitative stimulators, rather than by fluctuations in the chemical concentrations of certain ionic species.  

\subsubsection{The Leaky-Integrate and Fire (LIF) Neuron Model}

\qquad An alternative model with which to begin this analysis is the Leaky-Integrate and Fire (LIF) Neuron model.  Since the activity of the neurons in the synthetic network is not likely to be very similar to the electrodynamics associated with neurons \textit{in vivo}, the LIF Neuron model may be a more appropriate starting point for this analysis. This is because the mathematical expressions used in this model were not derived based on the physiology of particular neurons, but instead were derived in terms of general circuit elements (unlike the Morris-Lecar model which was derived specifically in terms of Ca$^{+2}$ and K$^+$ conductances).  This provides the LIF Neuron model with some flexibility in order to account for the fact that the electrical stimulation of these neurons is purely artificial.  This model can be expressed as follows:

\begin{equation}
 a_i(x) ~ = ~ G_i[J_i(x)]
 \end{equation}

\begin{equation} 
a_i(x) ~ = \left\{ \begin{array}{ll}
 \frac{1}{\tau _i^{\textrm{ref}} - \tau _i^{RC} \, \textrm{ln} \left( 1 - \frac{J_i^{\textrm{threshhold}}}{\alpha _i x + J_i^{\textrm{bias}} } \right)}  & \textrm{if} ~~ \alpha_i x + J_i^{\textrm{bias}} > J_i^{\textrm{threshold}}\\
0 & \textrm{otherwise}\\
\end{array} \right.
\end{equation}

\begin{equation}
	V(t) = J_M R \left( 1 - e^{t/\tau ^{RC}} \right)
\end{equation}

\noindent
where spikes occur when $\alpha_i x + J_i^{\textrm{bias}} > J_i^{\textrm{threshold}}$, and are approximated as infinitesimally small intervals of infinite voltage (i.e. Dirac-Delta functions). 

Accordingly, the spike train can be represented with respect to time as a sum of Dirac-Delta functions, each representing the occurrence of a spike at time $t_n$:

\begin{displaymath}
\sum_n \delta (t-t_n) 
\end{displaymath}

Once again a complete list describing the identities of the above symbols is provided in the appendix.  


In either the case of the LIF Neuron model or the ML model, we must tailor the model to characterize the \textit{Lymnaea stagnalis} particularly - which will involve experimenting with the free parameters in the model in order to synchronize the virtual behaviour with the results presented in \cite{2005Merz}.  A comprehensive model of \textit{Aplysia} neurons reported in Ref. \cite{1995Butera} should provide a good starting point for investigating the \textit{Lymnaea} neurons (no reports with such detailed models have been found for \textit{Lymnaea} yet).

\subsubsection{Extentions of the ML and LIF models}

\qquad  To complement the above preliminary models, we must add at least two more degrees of sophistication - we must modify these preliminary models (which only consider ion flow from the soma of one neuron to the soma of another neuron) to absorb the fact that the ion flow considered here occurs between neurites rather than cell bodies, and we must determine whether or not it is necessary to incorporate spatial dependence in the system of differential equations (DEs).  To address spatial dependence, we can start with the partial differential equation of cable theory for dendritic neurons\cite{2001Rall}:

\begin{equation}
\lambda ^2 (\frac{\partial ^2V}{\partial x^2} ) - V = \tau ( \frac{\partial V}{\partial t})
\end{equation}

\noindent and examine how dependent the steady-state voltage value is on small perturbations in the position values.  This will provide some insight with regards to whether or not it is necessary to add spatial dependence in the model.  

In lieu of this partial differential equation, incorporating an explicit time-delay in our model may be sufficient (although this time-delay may need to be a function of the length of a particular neurite, in the case where many of the neurites in the network are of different lengths).    

\subsection{Predictions \& Comparisons with Physical Realizations}

\qquad Once an appropriate modeling framework has been identified, we can test its performance for the simplest possible network: a pair of interconnected neurons.  The results of the simulation will then be compared with the results presented in Fig. 3 and modified according to how well the results match, until we are satisfied with the model. This stage of the analysis may involve adding noise terms to the system of DEs in order to account for the ubiquitous effects of the environment.

  \begin{figure}[h!]
  \centering
  \includegraphics[width = \textwidth , height = 0.25 \textheight] {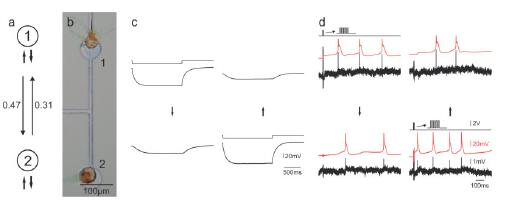} 
  \caption{Experimental results for a simple synthetic network of two mutually synapsing neurons stimulated by  capacitative stimulators.  This figure was reproduced with the kind permission of the publisher and authors of Ref. \cite{2005Merz}.  Copyright Wiley-VCH Verlag GmbH \& Co. KGaA.} 
\end{figure}

 Subsequently, the model will be tested against the results provided for a network of \textit{four} interconnected neurons: Fig. 4.  The simulation should also be able to predict reasonable values (between 0.02 and 0.80)\cite{2000Prinz} for the coupling coefficients that the experiments were unable to determine (the dotted lines shown in Fig. 4a).  
 
 
 Once this model is able to successfully reproduce the experimental results for the two-neuron and four-neuron systems, I will attempt to apply the model to a much larger (virtual) network of neurons - one capable of performing useful computations such as addition of scalar signals or a neural integrator\cite{EliasmithBook}.

  Theoretically, we should be able to predict the electrophysiological response in a neuron at the end of a chain, based on a known stimulus to a neuron at the beginning of the chain, and the results of such simulations will provide us with insight regarding the mechanisms by which we can manipulate these networks to perform computations.  We will then be ready to numerically simulate neurocomputational experiments, and attempt to implement simple logical operations such as AND gates, NOT gates, FANOUT gates, and combinations of the above.  
  
Since all computations (according to the Church-Turing thesis) can be performed as a combination of the above logic gates, the simulations made here will demonstrate that synthetic neuronal networks do in fact have the potential to form the basis of a scalable biological computing device.

\begin{figure}[h!]
  \centering
  \includegraphics[width = \textwidth, height=0.5\textwidth] {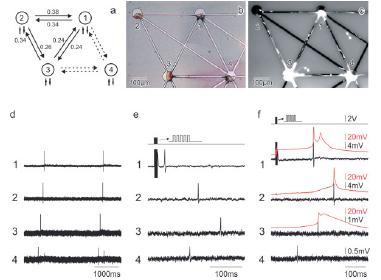} 
  \caption{Experimental results for a simple synthetic network of four mutually synapsing neurons stimulated by  capacitative stimulators.  This figure was reproduced with the kind permission of the publisher and authors of Ref. \cite{2005Merz}.  Copyright Wiley-VCH Verlag GmbH \& Co. KGaA.}
\end{figure}

\newpage 



\newpage

\appendix

\section{Appendix}

\subsection{Morris-Lecar Model Symbol Identifications}

$C$ = Membrane Capacitance \\
$V$ = Membrane Potential \\
$g_{Ca}$ = $Ca^{+2}$ ion conductance \\
$V_{Ca}$ = Nernst Potential for calcium ions  \\
$g_k$ = $K^+$ ion conductance \\
$V_K$ = Nernst Potential for potassium ions \\
$g_L$ = Conductance for a leak of ions \\
$V_L$ = Nernst Potential for the leak of ions \\
$I_{app}$ = Applied current \\
$\tau$ = Time constant for action potential \\
$w$ = Recovery variable \\
$\phi$ = Time constant for recovery process \\
$v_i$ = Free variables (fitted to different systems of interest)

\subsection{Leaky-Integrate and Fire Neuron Model Symbol Identifications}

$a$ = Firing rate of neuron \\
$G$ = A non-linear function of the input current \\
$J$ = Input current \\
$\tau^{\textrm{ref}}$ = Refractory period \\
$\tau^{\textrm{RC}}$ = Characteristic time \\
$J^{\textrm{threshold}}$ = Threshold current (current required for firing) \\
$J^{\textrm{bias}}$ = Noise term to represent the current from the environment \\
$V$ = Membrane potential \\
$J_M$ = Current through membrane \\
$R$ = Resistance of membrane \\
$C$ = Capacitance of membrane \\
$t$ = Time \\
$x$ = Analog signal \\
$\alpha$ = Gain factor

\subsection{The Partial Differential Equation of Cable Theory}

$\lambda$ = Characteristic length \\ 
$\tau$ = Characteristic time \\
$V$ = Membrane potential \\
$x$ = Spatial coordinate \\
$t$ = Time coordinate

\end{document}